\documentclass[12pt]{article}
\usepackage[english]{babel}%
\usepackage{graphicx}
\usepackage{amsfonts}

\def\qed{{\bf \hfill $\Box$}\endtrivlist}

\def\rank{\mathbf{rank}}
\def\tr{\mathbf{tr}}
\def\diag{\mathbf{diag}}

\begin{document}

\newtheorem{theorem}{Theorem}
\newtheorem{cor}{Corollary}
\newtheorem{example}{Example}

\renewcommand{\thesection}{\arabic{section}}
\renewcommand{\thetheorem}{\arabic{section}.\arabic{theorem}}
\renewcommand{\theequation}{\arabic{section}.\arabic{equation}}
\renewcommand{\theexample}{\arabic{section}.\arabic{example}}
\renewcommand{\thecor}{\arabic{section}.\arabic{cor}}

\title{Distance Evaluation to the Set of Defective Matrices}

\author{%
{Alexei Yu. Uteshev\footnote{The corresponding author}, Elizaveta A. Kalinina, Marina V. Goncharova }%
\vspace{1.6mm}\\
\fontsize{10}{10}\selectfont\itshape
% separate superscript on following line from affiliation using narrow space
%$^{\#}$\,
St.\,Petersburg State University\\
\fontsize{10}{10}\selectfont\itshape Faculty of Applied Mathematics \\
\fontsize{10}{10}\selectfont\itshape	St.\,Petersburg, Russia\\
\fontsize{9}{9}\selectfont\ttfamily\upshape
$^{1}$\,\{alexeiuteshev,ekalinina,marina.yashina\}@gmail.com\\
%$^{2}$\, ekalinina,@gmail.com
\vspace{1.2mm}\\
\fontsize{10}{10}\selectfont\rmfamily\itshape
\fontsize{9}{9}\selectfont\ttfamily\upshape
}

\maketitle
\thispagestyle{empty}

\begin{abstract}
%\boldmath
We treat the problem of the Frobenius distance evaluation from a given matrix $ A \in \mathbb R^{n\times n} $ with distinct eigenvalues to the manifold  of matrices with multiple eigenvalues. On restricting considerations to the  
rank $ 1 $ real perturbation matrices, we prove that the distance in question equals $ \sqrt{z_{\ast}} $ where 
$ z_{\ast} $ is a  positive (generically, the least positive) zero of the algebraic equation
$$
\mathcal F(z) = 0, \ \mbox{where} \ 
\mathcal F(z):= \mathcal D_{\lambda} \left( 
\det \left[ (\lambda I - A)(\lambda I - A^{\top})-z I_n \right] \right)/z^n
$$
and $ \mathcal D_{\lambda} $ stands for the 
discriminant of the polynomial treated with respect to $\lambda $. In the framework of this approach we also provide the procedure for finding the nearest to $ A $ matrix with multiple eigenvalue. Generalization of the problem to the case of complex perturbations is also discussed. Several examples are presented clarifying the computational aspects of the approach.

\textbf{Keywords:} Wilkinson's problem, defective matrix, multiple eigenvalues, distance equation

\textbf{MSC 2010:} 68W30, 15A18 , 12D10, 58C40

\end{abstract}

%\begin{IEEEkeywords}
%\keywords{Polynomial  interpolation, rational interpolation, Hankel matrices and polynomials, error correction, Berlekamp---Massey algorithm, resultant interpolation
%\end{IEEEkeywords}

% \fi
%\IEEEpeerreviewmaketitle

\section{Introduction}
\label{SIntro}

\setcounter{equation}{0}
\setcounter{theorem}{0}
\setcounter{example}{0}

The origin of the problem of finding the distance from a matrix $ A \in \mathbb C^{n\times n} $
to the set $ \mathbb D $ of matrices with multiple eigenvalues can be traced back to Wilkinson~\cite{Wilkinson} who posed it in relation to the sensitivity analysis of matrix eigenvalues. The desired distance  further will be treated  with respect to either the $ 2 $-norm or to the Frobenius norm in $ \mathbb C^{n\times n} $ and will be 
denoted $d(A, \mathbb D)$. It is usually referred to as the {\it Wilkinson
distance} of $A$ \cite{AlamBora,Mengi}. Alternatively, 
$d(A, \mathbb D)$ can be defined as the 
$\inf \|A-B \| $ where $ B $ belongs to the subset of defective matrices, i.e. those possessing at least one eigenvalue 
whose geometric multiplicity  is less than its algebraic multiplicity.

%In the present paper we refer to the defective matrix as the %one possessing a multiple eigenvalue (regardless its geometric %multiplicity).

%In other words, a defective matrix has at least one defective %eigenvalue~\cite{AlamBoraByersOverton}

Starting from  Wilkinson's works~\cite{Wilkinson2,Wilkinson1,Wilkinson3}, 
the problem of evaluation of $d(A, \mathbb D)$ has been
studied intensively in~\cite{AlamBora,Demmel,Demmel1,LipEdel,Malyshev,Ruhe}. The most recent result is presented in the work~\cite{AkFrSp}. 
We briefly trace the developed approaches. Most of them are in the framework of singular value analysis of appropriate parameter dependent matrices.

The following theorem gives the min-max representation of $d(A,\mathbb D)$ obtained by Malyshev~\cite{Malyshev}.

\begin{theorem}\label{Malyshev} Let $A\in \mathbb C^{n\times n} $.
Let the singular values of the matrix
\begin{equation}
\left[\begin{array}{cc}
A-\lambda I_n&\gamma I_n\\
\mathbb{O}_{n\times n}&A-\lambda I_n \end{array}\right]
\label{MatrM}
\end{equation}
be ordered like
$\sigma_1(\lambda,\gamma)\ge\sigma_2(\lambda,\gamma)\ge\ldots\ge\sigma_{2n}(\lambda,\gamma)
\ge 0 $. Then the  $2$-norm distance  $ d(A,\mathbb D) $ can be evaluated as 
$$d(A,\mathbb D)=\min_{\lambda\in\mathbb{C}}\max_{\gamma\ge0} \sigma_{2n-1}(\lambda,\gamma)\, .$$
\end{theorem}

The straight computation of this distance is quite difficult, so to find this distance, in many works the notion of pseudospectra~\cite{TrEmb} is used.

{\bf Definition.} For both the 2-norm and the Frobenius norm, the $\varepsilon$-pseudospectra of a matrix $A$
is
$$\Lambda_{\varepsilon}(A)=\{ \sigma_{\min}<\varepsilon \}$$
where $\varepsilon>0$ and $\sigma_{\min}$ stands for the smallest
singular value of the matrix $A-zI$.

Equivalently,
$$\Lambda_{\varepsilon}(A)=\{ z\in\mathbb{C} | \det(A+E-zI)=0, \mbox{ \rm for some } E\in\mathbb{C}^{n\times n} \mbox{ \rm with } \|E \|<\varepsilon\}\, .$$
If $\Lambda_{\varepsilon}$ has $n$ components, then $A+E$ has $n$
distinct eigenvalues for all perturbations
$E\in\mathbb{C}^{n\times n}$ and hence $A+E$ is not defective.

In subsequent papers, the pseudospectra approach is used to find the distance to the nearest defective matrix. 

In~\cite{LipEdel}, a geometric solution to the problem of finding $
d(A,\mathbb D) $ in Frobenius norm is given. The nearest defective
matrix is related to the critical points of 
the minimal singular value $\sigma_{\min}(x,y)$ of the matrix $A-(x+\mathbf i y)I$ that could
be obtained by examination of pseudospectra of $A$. For an
approximation of a multiple eigenvalue of the nearest defective
matrix, the averaging heuristic by Puiseux series is proposed. Also
an iterative method for finding this eigenvalue
together with the minimal perturbation is presented.

In~\cite{AlamBora}, it is proposed to find the smallest
perturbation $E$ such that the components of the pseudospectra of
$A+E$ coalesce. The problem is reformulated as follows. One needs to
find
$z\in\mathbb{C},\varepsilon\in\mathbb{R},\varepsilon>0$ and $U,V\in\mathbb{C}^n $,\
such that\footnote{Hereinafter $ \mbox{}^{\top} $ stands for the transpose while $ \mbox{}^{\mathsf H} $ stands for the Hermitian transpose.}
\begin{equation}\label{AlBor}
(A-zI)V-\varepsilon U=\mathbb O_{n\times 1},\varepsilon V-(A-zI)^{\mathsf H} U=\mathbb O_{n\times 1},\, U^{\mathsf H} V=0 \, .
\end{equation}
The algorithm to solve the system of equations presented in this
work is rather expensive because it requires the repeated
calculation of pseudospectra. Also any condition of coalescence of
two pseudospectral curves is necessary.

In~\cite{AlamBoraByersOverton}, a new computational approach to approximating the nearest defective matrix by a variant of Newton's method is  suggested. 

The Implicit Determinant Method based on standard Newton's method
is used to solve the system~(\ref{AlBor}) in~\cite{AkFrSp}.

There are several works considering generalizations of Wilkinson's problem for the cases of prescribed eigenvalues or their  multiplicities ~\cite{ArmGraVel,Gracia,Lippert,KokLogKar,Mengi},
 and matrix pencils~\cite{AhmadAlam}.

The approaches developed in the above cited papers could be characterized as related to the Numerical Linear Algebra. The present paper aims at solving the stated problem for the case of Frobenius norm within the framework of symbolic computation approach. Namely, we reduce the problem to that of the univariate polynomial equation solving. As a matter of fact, the manifold $\mathbb D$ of  matrices with multiple eigenvalues in the $\mathbb R^{n^2} $ space of their entries is an algebraic one, i.e. it is represented by a multivariate polynomial equation. If we slightly modify  Wilkinson's problem to that of finding $ d^2(A,\mathbb D) $, then the constrained optimization problem becomes an algebraic one in the sense that both the objective function and the constraint be polynomials.  Application of the Lagrange multipliers method reduces the problem to that of system of algebraic equations solving. The latter can be resolved, at least in theory, via the analytical procedure of elimination of variables consisting in the multivariate resultant computation or the Gr\"obner basis construction. Application of these procedures to the system of equations of the treated problem, complemented with $z-d^2(A,\mathbb D)=0$, results in a univariate equation $\mathcal F(z)=0$ whose zero set contains all the  critical values of the squared distance function. This equation will be further referred to as the {\bf distance equation} and its computation is the priority of the present paper. 

This approach has been developed in \cite{Kalinina_Uteshev_CASC22}.
Unfortunately, soon after that publication, a significant gap in reasoning was discovered. It was assumed that the value $ d(A, \mathbb D)$ could be provided by only the rank $1$ perturbation matrix $ E_{\ast} $ and that the nearest to $ A $ matrix $B_{\ast}=A+E_{\ast}$ in $ \mathbb D $ might possess only a double real eigenvalue.
In Section \ref{SExcep}, an example of the order $ 4 $ matrix $ A $ is given where the nearest  in $ \mathbb D $  matrix
  possesses a pair of double complex-conjugate eigenvalues. 
As yet we failed to manage this scenario for the general statement of the problem; neither do we able to claim that it is a {\it zero probability event}. 

We confine ourselves here to considering the case where the critical values of $d^2(A,\mathbb D) $ are provided only by the rank $ 1 $ perturbation matrices. For this case,  
the practical implementations of the elimination of variables procedure mentioned above  can be reduced to just only two bivariate equations. 
One of these equations follows quite naturally from the developed in \cite{Kalinina_Uteshev_CASC22} approach. This is 
$$
\Phi(\lambda,z)=0 \ \mbox{where} \ 
\Phi(\lambda,z):=
\det \left[ (\lambda I - A)(\lambda I - A^{\top})-z I_n \right]
 \, .
$$
The more difficulties causes the deduction of the second equation. It happens to be $$
\partial \Phi(\lambda,z)/ \partial \lambda =0 \, . 
$$
To obtain the distance equation, it is then sufficient to eliminate the variable $\lambda $ from the obtained system. This can be managed with the aid of {\bf discriminant} computation, i.e. the function of the coefficients of a polynomial responsible for the existence of a multiple zero for this polynomial.
We recall some basic features of this function in Section \ref{SPrelim}.

In Section \ref{SDisEq}, we prove the main result of the paper, namely that the value $ d^2(A,\mathbb D) $ is in the set of non-negative zeros of the distance equation. If $A \not\in \mathbb D $ then generically $ d^2(A,\mathbb D) $ equals the least positive zero $ z_{\ast} $ of this equation. We also detail here the structure of the matrix $ B_{\ast} $ nearest  to $ A $ in $\mathbb D$. It appears that the multiple eigenvalue of  $ B_{\ast} $ coincide with the multiple zero of the polynomial $ \Phi(\lambda,z_{\ast}) $.

In Section \ref{SExamples}, computational aspects of the proposed approach are discussed via solving the  problem for the two families of matrices treated in  the literature. 

In Section \ref{GenCase}, we address to the generalization of Wilkinson's problem to the case of complex perturbations. Here the results are presented in a very concise manner with the potential intention of returning to them in future articles.

{\bf Notation} is kept to correlate with \cite{Kalinina_Uteshev_CASC22}. For a matrix $A \in \mathbb R^{n\times n} $,
$f_A(\lambda) $ denotes its characteristic polynomial,
%$\adj (A)$ stands for its adjoint matrix,
$ d(A, \mathbb D) $ denotes the distance from $ A $ to the set $ \mathbb D $ of matrices possessing a multiple eigenvalue.  $ E_{\ast} $ and $ B_{\ast} = A+ E_{\ast} $ stand for, respectively, the (minimal) perturbation matrix and the nearest to $ A $  matrix in $ \mathbb D $ (i.e. $d(A,\mathbb D)=\|A- B_{\ast}\|$); we then term by $\lambda_{\ast}$ the multiple eigenvalue of $B_{\ast}$. $ I $ (or $ I_n$) denotes the identity matrix (of the corresponding order). $\mathcal D$ (or $\mathcal D_{\lambda} $) denotes the discriminant of a polynomial (with subscript indicating the variable).

\textbf{Remark.} All the computations were performed in CAS Maple 15.0 with those approximate done within the accuracy $ 10^{-40} $. In the paper they are presented rounded to $ 10^{-6} $.

\section{Algebraic Preliminaries}
\label{SPrelim}

\setcounter{equation}{0}
\setcounter{theorem}{0}
\setcounter{example}{0}

It is well-known that in the $(N+1)$-dimensional space of the polynomial 
\begin{equation}
F(x)=a_0x^N + a_1x^{N-1} +\dots+a_N \in \mathbb C[x], \, a_0\ne 0, N \ge 2
\label{polyF}
\end{equation}
coefficients, the manifold of polynomials with multiple zeros 
is given by the equation
\begin{equation}
D(a_0,a_1,\dots,a_N)=0 \quad \mbox{where} \ 
D:=\mathcal D_x(F(x))
\label{DiscrFl}
\end{equation}
denotes the {\bf discriminant} of the polynomial. Discriminant is formally defined as a symmetric function of the zeros $\{\lambda_1,\dots, \lambda_N \} $ of the polynomial $ F(x)$
\begin{equation}
D_x(F(x))= a_0^{2N-2} \prod_{1\le j < k \le N} (\lambda_k - \lambda_j)^2 \, .
\label{DiscDef}
\end{equation}
This representation gives rise to further transformation of the discriminant into the homogeneous polynomial $ D(a_0,a_1,\dots,a_N) $ of the order $2N-2$ with respect to the coefficients of $ F(x) $. Such a transformation  can be implemented through a preliminary representation of discriminant in an appropriate determinantal form. We will follow the approach based on the Hankel matrix formalism \cite{UteshevCherkasov}.

For this aim, find first the {\it Newton sums} $ s_{0},s_1,\dots,s_{2N-2} $ of the polynomial $ F(x) $ with the aid of recursive formulas
$$s_0=N,\ s_1=-a_1/a_0,\ $$
\begin{equation}
s_k=\left\{\begin{array}{lr}
-(a_1s_{k-1}+a_2s_{k-2}+\dots+a_{k-1}s_1+a_kk)/a_0,
& \mbox{ if} \ k\le N ,\\
-(a_1s_{k-1}+a_2s_{k-2}+\dots+a_Ns_{k-N})/a_0,
& \mbox{if} \ k > N ,
\end{array}
\right.
\label{Newton_sums}
\end{equation}
and compose the Hankel matrix
$$
S=\left[s_{j+k} \right]_{j,k=0}^{N-1} =
\left[\begin{array}{llllll}
s_0 &s_1&s_2&\dots&s_{N-2}& s_{N-1}\\
s_1 &s_2&s_3&\dots&s_{N-1}& s_{N}\\
s_2 &s_3&s_4&\dots&s_{N}& s_{N+1}\\
\dots& & &&& \dots\\
s_{N-1} &s_N&s_{N+1}&\dots &s_{2N-3}&s_{2N-2}
\end{array}\right]_{N\times N} \ .
$$
Denote by $ S_{1},\dots, S_N=\det S $ its leading principal minors.

\begin{theorem} \label{ThDiscrHankel}  One has
\begin{equation}
{\mathcal D}(F)=a_0^{2N-2} S_N \, .
\label{SnDisc}
\end{equation}
The condition 
$$S_N=0, \dots, S_{N-k+1}=0, S_{N-k}\ne 0   $$
is the necessary and sufficient for the polynomial $ F(x) $ to possess $ k $ 
common zeros with $ F^{\prime} (x)$.
In particular, if $ S_N=0, S_{N-1}\ne 0 $, then $ F(x) $ possesses a unique multiple zero and the multiplicity of this zero equals $ 2 $. This zero can be computed via the formula
\begin{equation}
\lambda = 
s_1-\frac{1}{S_{N-1}}
\left|
\begin{array}{lllll} 
s_0 & s_1 & \dots & s_{N-3} & s_{N-1} \\ 
s_1 & s_2 & \dots & s_{N-2} & s_{N} \\ 
\vdots & & & & \vdots \\
s_{N-2} & s_{N-1} & \dots & s_{2N-1} & s_{2N-3}
\end{array} \right|  \, . 
\label{double_zero}
\end{equation}
The determinant in the right-hand side is constructed by deleting the last row and the last but one column in 
$ \det S $.
\end{theorem}

Consequently, the set $ \mathbb D $ of matrices with multiple eigenvalues is given by the equation
$$ \mathcal D_{\lambda} \left(\det (\lambda I-B) \right) =0 \, .
$$

For the case of polynomials with real coefficients, the sequence of leading principal minors of the matrix $ S $ permits one to establish the exact number of real zeros for $ F(x) $ \cite{Gantmacher}.

\begin{theorem}[Jacobi]\label{ThJacobi} Let 
$$S_N=0, \dots, S_{N-k+1}=0, S_{N-k}\ne 0,\dots,S_1 \ne 0   $$
Then the number of distinct pairs of complex-conjugate zeros for $ F(x) \in \mathbb R[x] $ equals
$$
\mathcal V(1,S_1,\dots, S_{N-k})  
$$
where $ \mathcal V $ denotes the number of variations of sign in the given sequence.
\end{theorem}

In the space $ \mathbb R^{N+1}$ of polynomials (\ref{polyF}) with real coefficients, the discriminant manifold (\ref{DiscrFl}) separates the domains of vectors providing the coefficients of polynomials with the same number of real zeros. 

The last comment of the present section relates to application of discriminant to one problem from Elimination Theory. Consider a bivariate polynomial $ F(x,y)\in \mathbb R[x,y], \deg F \ge 2 $. The discriminant furnishes the tool for eliminating the variable $ x $ from the system of equations
\begin{equation}
F(x,y)=0, {\partial F(x,y)}/ {\partial x}=0 \, .
\label{fdf}    
\end{equation}    

Namely, if $(x_0,y_0) $ is a solution to the system (\ref{fdf}), then $ y_0 $  is necessarily a zero of the algebraic univariate equation
$$
\mathcal Y(y)=0 \ \mbox{ where } \ \mathcal Y(y):=\mathcal D_x(F(x,y)) \, .
$$
The reverse statement is subject to an extra assumption. If $ y_1\in \mathbb C $ is a zero for $ \mathcal Y(y)$, then there exists a multiple zero for the polynomial $ F(x,y_1)$.  Under the assumption that $ y_1 $ is a simple zero for $ \mathcal Y(y) $,
$ x_1 $ is a  unique multiple zero and its  multiplicity equals $ 2 $. Then it can be expressed as a rational function of $  y_1 $  using the result of Theorem \ref{ThDiscrHankel}. These considerations are valid for all the solutions of the system (\ref{fdf}) provided that $\mathcal D_y(\mathcal Y(y))\ne 0$.

\section{Distance Equation}
\label{SDisEq}

\setcounter{equation}{0}
\setcounter{theorem}{0}
\setcounter{example}{0}
\setcounter{cor}{0}

In terms of the discriminant manifold referred to in the previous section, the problem of evaluation of $d^2 (A,\mathbb D) $ is equivalent to that of constrained optimization
$$
\min \|B-A\|^2 \quad \mbox{subject to} \ \mathcal D_{\lambda}(f_B(\lambda))=0,\ B\in \mathbb R^{n\times n} \, .
$$
Here the constraint is an algebraic equation with respect to the entries of the matrix $B$.
Traditional application of the Lagrange multipliers method reduces the problem to that of solving a system of $ n^2+1 $ nonlinear algebraic equations. Under the additional assumption the matrix $ B_{\ast} \in \mathbb R^{n\times n} $ providing a solution to this system possesses only one multiple eigenvalue and its multiplicity equals $ 2 $, it is possible to  reduce the number of variables in the constrained optimization approach. The following result is presented in \cite{Kalinina_Uteshev_CASC22}:

\begin{theorem}
The value $d^2(A,\mathbb D) $ belongs to the set of critical values of the objective function
\begin{equation}
G(U):=U^{\top}A A^{\top} U - \left( U^{\top}AU \right)^2
\label{eqG}
\end{equation}
for the constrained optimization problem under constraints 
\begin{equation}
U^{\top}U=1,\ U \in \mathbb R^n \, . 
\label{eqUTU}
\end{equation}
If $ U_{\ast} $ be the point providing $ d^2(A,\mathbb D) $, then the perturbation can be computed as
$$
E_{\ast}=U_{\ast} U_{\ast}^{\top} (\kappa I-A) \quad \mbox{ where} \ \kappa:= U_{\ast}^{\top} A U_{\ast} \, .
$$
\end{theorem}

The new optimization problem still have significant number of variables. We aim to eliminate all of them but introduce an extra one responsible for the critical {\it values} of the objective function. 

Stationary points of the function
(\ref{eqG}) under the constraints
(\ref{eqUTU}) can be found via Lagrange  method applied to the function $ G(U)- \mu (U^{\top}U-1)$. This results into the system
\begin{equation}
AA^{\top}U-(U^{\top} A U)(A+A^{\top})U-\mu U = \mathbb O_{n\times 1} \, .
\label{eqMain}
\end{equation}
Denote
\begin{equation}
\lambda:=U^{\top} A U \, .
\label{eqLambda}
\end{equation}
 Then the equation (\ref{eqMain}) has a nontrivial solution with respect to  $ U $ if and only if
 \begin{equation}
 \det(AA^{\top}-\lambda (A+A^{\top})-\mu I)=0 \, . 
 \label{eqdet}
 \end{equation}
 Under this condition, multiplication of 
  (\ref{eqMain}) by $ U^{\top}$ yields
 $$
 U^{\top}AA^{\top}U=2\lambda^2+\mu \, .
 $$
 Wherefrom it follows that the critical values of the objective function
  (\ref{eqG}) are given by 
 $$
 z=\lambda^2+\mu \, .
 $$
 Substitution this into (\ref{eqdet}) results in the equation connecting  $ z $ and $\lambda$:
\begin{equation}
 \Phi(\lambda,z)=0
 \label{eqPhi0}
 \end{equation}
 where
\begin{equation}
\Phi(\lambda,z):=
\det \left[ A A^{\top}- \lambda (A+A^{\top})+(\lambda^2-z) I \right]
\label{eqPhi00}
\end{equation}
\begin{equation}
=\det  \left[ (\lambda I - A)(\lambda I - A)^{\top}-z I \right]
\label{eqPhi}
\end{equation}
Zeros $ z_1,\dots, z_n $ of the polynomial $\Phi(\lambda,z)$  with respect to the variable $ z $ are evidently real since they are the squares of the singular values for the matrix $ \lambda I-A $.

Our further task is to deduce an extra equation connecting 
$\lambda$ and $ z$.

\begin{theorem} The value $d^2(A,\mathbb D) $ belongs to the set of non-negative zeros of 
the polynomial
\begin{equation}
\mathcal F(z)\equiv \mathcal D_{\lambda}(\Phi(\lambda,z))/z^n \, .
\label{eqFz}
\end{equation}
\end{theorem}

\textbf{Proof.} Under the condition (\ref{eqPhi0}), there exists  a nontrivial solution  for (\ref{eqMain}) with respect to the column  $ U $
\begin{equation}
 (\lambda I-A)(\lambda I-A)^{\top}U=z\, U  \, .
 \label{eqGen}  
\end{equation}
This equality means that $ U $ is the right 
singular vector for the matrix $\lambda I - A $ corresponding to the singular value $\sqrt{z}$. The corresponding left singular vector for that matrix can be found from the equality
\begin{equation}
 \sqrt{z}V:=(\lambda I - A)^{\top}U \, .
 \label{eqV}
\end{equation}
Dual relationship is valid for $ U $:
\begin{equation}
 \sqrt{z}U=(\lambda I - A)V \, .
 \label{eqU}
\end{equation}
From the conditions (\ref{eqUTU}) and (\ref{eqLambda})
%\begin{equation}
%U^{\top}U=1, \ U^{\top}AU=\lambda \, .
%\label{eq_cond}
%\end{equation}
it follows that
\begin{equation}
 U^{\top}(\lambda I - A)U=0 \, .
 \label{eq1}
\end{equation}
Multiply (\ref{eqU}) from the left by $U^{\top}$. From (\ref{eqUTU}), it follows that
\begin{equation}
 \sqrt{z}=U^{\top}(\lambda I - A)V \, .
 \label{eqUTV}
\end{equation}
Multiply (\ref{eqV}) from the left by $V^{\top}$ and utilize (\ref{eqUTV}):
$$
\sqrt{z}V^{\top}V=V^{\top}(\lambda I - A)^{\top}U=\sqrt{z} \, .
$$
Wherefrom the two alternatives follow
$$
V^{\top}V=1 \quad \mbox{or} \quad \sqrt{z}=0 \, .
$$
Similarly, multiplication of (\ref{eqV}) from the left by $U^{\top}$ and further application of (\ref{eq1}) yields
$$
\sqrt{z}U^{\top}V=0 \, .
$$
This also leads to two alternatives:
$$
U^{\top}V=0 \quad \mbox{or} \quad \sqrt{z}=0 \, .
$$
Ignore the case $\sqrt{z}=0 $. 
\begin{equation}
V^{\top}V=1,\  U^{\top}V=0 \, .
\label{eqVTUT}
\end{equation}
Consider the equation (\ref{eqUTV}) as a definition of the $\sqrt{z}$ as the function of $ \lambda $. Differentiate this relation with respect to $ \lambda $:
$$
\frac{d\, \sqrt{z}}{d\, \lambda}=U^{\top}V+\frac{d\, U^{\top}}{d\, \lambda} (\lambda I - A)V+U^{\top} 
(\lambda I - A)  \frac{d\, V}{d\, \lambda} \, .
$$
With the aid of (\ref{eqV}) and (\ref{eqU}) transform this into
$$
U^{\top}V+\sqrt{z} \left[ \frac{d\, U^{\top}}{d\, \lambda} U + V^{\top}  \frac{d\, V}{d\, \lambda}  \right] \, .
$$
Due to (\ref{eqUTU}) and (\ref{eqVTUT}), we arrive at 
\begin{equation}
\frac{d\, \sqrt{z}}{d\, \lambda} = 0 \, .
\label{eqdsqrtz}
\end{equation}

Equation (\ref{eqPhi0}) defines implicit function $z(\lambda) $. Differentiation of the identity $\Phi(\lambda,z(\lambda))\equiv 0$ with respect to $\lambda $ yields the identity
$$
\Phi^{\prime}_{\lambda}(\lambda,z)+\Phi^{\prime}_{z}(\lambda,z) \frac{d\, z}{d\, \lambda}\equiv 0 \, .
$$
Under the condition (\ref{eqdsqrtz}), the variables $ \lambda $ and $ z $ are linked by an extra relationship
\begin{equation}
\Phi^{\prime}_{\lambda}(\lambda,z)=0 \, .
\label{eqdifPhi}
\end{equation}
Together with (\ref{eqPhi0}),  the deduced condition composes the system of algebraic equations
\begin{equation}
\Phi (\lambda,z)=0,\  \Phi^{\prime}_{\lambda}(\lambda,z)=0  \, .
\label{eqsysT}    
\end{equation}
According with the results of Section \ref{SPrelim}, elimination of $\lambda $ from this system can be implemented with the aid of the discriminant computation, i.e. the variable $z $ should  satisfy the equation
$$
\mathcal D_{\lambda} (\Phi (\lambda,z))=0 \, .
$$
To prove the validity of (\ref{eqFz}), it is necessary to additionally confirm that the left-hand side of the last equation is divisible by $ z^n $. This is indeed the case, since the polynomial $\Phi(\lambda,0)$ possesses $ n $ multiple zeros coinciding with the eigenvalues of the matrix $ A $. \qed

With $ \mathcal F(z) $ given by (\ref{eqFz}), the {\bf distance equation} $ \mathcal F(z) =0 $ is now well-defined and in Section \ref{SProper} we discuss some of related features and computational aspects. 

To conclude the present section, we have to detail the properties of the $\lambda$-component for the solution of the system (\ref{eqsysT}).
Let the polynomial $\mathcal F(z)$ defined by (\ref{eqFz}) possess a positive real zero $ z_0 $ and this zero be simple. Then the polynomial $ \Phi (\lambda,z_0) $ has a unique multiple zero and multiplicity of this zero equals $ 2 $. We denote by $\lambda_0$. It is evidently real and can be expressed as a rational function of $z_0$ via, for instance, formula (\ref{double_zero}). 

The less evident conclusion is as follows: this multiple zero coincides with the multiple eigenvalue of the matrix in $ \mathbb D $ providing the critical value $ z_0 $ for the function $ d^2(A,\mathbb D)$.

\begin{theorem} \label{Th3}
For any real solution $(\lambda_0,z_0)$ of the system~(\ref{eqsysT})  where
$z_0\ne 0$, there exists the rank $1$ perturbation $E_0$ such that 
$\|E_0\|=\sqrt{z_0}$ and the matrix $B_0=A+E_0$ possesses the multiple eigenvalue 
$\lambda_0$.
\end{theorem}

\textbf{Proof.} The number $\sqrt{z_0}$ is a singular value  for the matrix $ \lambda_0 I - A$. We intend to prove that the matrix from the theorem statement is defined by the formula
\begin{equation}
E_0:=\sqrt{z_0}U_0 V_0^{\top} \, , 
\label{eqEj}
\end{equation}
where $U_0$ and $V_0$ are respectively the left and the right singular vectors of the unit norm for the matrix
 $\lambda_0 I-A$ corresponding to $ \sqrt{z_0}$.

Indeed, the matrix $ B_0=A+E_0 $  has $\lambda_0$ as the eigenvalue corresponding to the eigenvector $V_0$: 
$$
B_0V_0=(A+E_0)V_0 
\stackrel{(\ref{eqEj})}= 
AV_0+\sqrt{z_0}U_0
\stackrel{(\ref{eqU})}
=AV_0+(\lambda_0I-A)V_0=\lambda_0V_0 \, .
$$
If $\rank (B_0-\lambda_0 I)<n-1$ then the theorem is proved. Assume that $\rank (B_0-\lambda_0 I)=n-1$. Let us prove the existence of a column $W$ such  that
$$ (B_0-\lambda_0 I)W=V_0 \, . $$
The necessary and sufficient condition for resolving this equation consists in the fulfillment of the equality
\begin{equation}
(B_0-\lambda_0I)(B-\lambda_0I)^{+}V_0=V_0 
\label{eqCond}
\end{equation}
where $\mbox{ }^{+}$ stands for the Moore-Penrose inverse of the matrix. It can be easily verified that
$$
(B_0-\lambda_0I)(B_0-\lambda_0I)^{+}=I-U_0U_0^{\top} 
$$
(by assumption, $\rank (B_0-\lambda_0 I)=n-1$), and the condition (\ref{eqCond}) is fulfilled:
$$(B_0-\lambda_0I)(B_0-\lambda_0I)^{+}V_0=
(I-U_0U_0^{\top})V_0
\stackrel{(\ref{eqVTUT})}=
V_0 \, .
$$

The columns $V_0$ and $ W $ are linearly independent. 
Indeed, if
$$\alpha V_0+ \beta W=\mathbb O_{n \times 1} \quad \mbox{for} \ \{\alpha, \beta\} \subset \mathbb R 
$$
then on multiplying this equality from the left by
$B_0-\lambda_0 I$ it follows that
 $\beta V_0 = \mathbb O_{n \times 1}$, and thus $ \beta=0$. But then $\alpha=0$ since $V_0$ is a nonzero column. 
 
Hence, 
 $$ (B_0-\lambda_0I)^2V_0=  \mathbb O, \ (B_0-\lambda_0I)^2W=  \mathbb O $$
 for the linear independent $V_0$ and $ W $. Consequently,
 $\rank (B_0-\lambda_0 I)^2\le n-2$ and this gives evidence that $\lambda_0 $ should be a multiple eigenvalue for $B_0$. \qed
 
\begin{cor}
If $ A \not\in \mathbb D $,  then
$$
d(A,\mathbb D) = \sqrt{z_{\ast}}  \, ,
$$
where $ z_{\ast} $ is the minimal positive zero of the polynomial (\ref{eqFz}) provided that this zero is not a multiple one. Minimal perturbation is evaluated by the formula
\begin{equation}
E_{\ast}=U_{\ast}U_{\ast}^{\top} (\lambda_{\ast}I-A) \, .
\label{min_perturb}
\end{equation}
Here $ \lambda_{\ast}$ is the multiple zero for the polynomial $ \Phi(\lambda,z_{\ast}) $ and $U_{\ast} \in \mathbb R^{n}, \|U_{\ast}\|=1 $ is the left singular vector of the matrix $\lambda_{\ast}I-A$ corresponding to the singular value $ \sqrt{z_{\ast}} $.
\end{cor}

The significance of condition for simplicity of the minimal positive zero $ z_{\ast} $ can be explained as follows. Since we are looking for only real perturbations, formula (\ref{min_perturb}) yields such a matrix if $ \lambda_{\ast}$ is real. For the matrices of the order $n\ge 4$, it might happen that the system (\ref{eqsysT}) possesses a solution $(z_{\ast}, \lambda_{\ast}) $ with an imaginary $ \lambda_{\ast} $ (we give an example of such a matrix in Section \ref{SExcep}). Then the system necessarily possesses the solution $ (z_{\ast},\overline{\lambda_{\ast}}) $. This implies (v. the last comment from Section \ref{SPrelim}) that  $ z_{\ast} $ should be a multiple zero for (\ref{eqFz}). Therefore, the condition for simplicity of $ z_{\ast} $is sufficient to prevent such an occasion. Formal verification of this condition can be replaced by a more general one relating the discriminant of $\mathcal F(z)$: 
$$ \mathcal D_z(\mathcal F(z))\ne 0 \, . $$

\section{Properties of the Distance Equation}
\label{SProper}

\setcounter{equation}{0}
\setcounter{theorem}{0}
\setcounter{example}{0}
\setcounter{cor}{0}

\begin{example} The distance equation for the matrix $A=[a_{jk} ]_{j,k=1}^2 $ is found in the form
$$
\mathcal F(z):=16\left[ (a_{11}-a_{22})^2+(a_{12}+a_{21})^2 \right]\cdot \left\{ \left[4z- \mathcal D (f_A(\lambda)) \right]^2-16(a_{12}-a_{21})^2z \right\}=0\, .
$$
Polynomial in braces has only real zeros with respect to $ z$ since its discriminant equals
$$
256(a_{12}-a_{21})^2\left[(a_{11}-a_{22})^2+ (a_{12}+a_{21})^2 \right] \ge 0 \, .
$$
\end{example}
\qed

Some terms in the canonical representation of the polynomial (\ref{eqPhi0}) can be explicitly expressed via the entries of the matrix $A$:
\begin{equation}
\Phi(\lambda,z)\equiv \lambda^{2n}- 2 \tr(A) \lambda^{2n-1} +(-nz+ \tr(AA^{\top})+p_2  )  \lambda^{2n-2} +\dots +
\det(AA^{\top}-zI) \, .
\label{Phi_can}
\end{equation}
Here $ p_2 $ is the coefficient of $\lambda^{n-2} $ in the characteristic polynomial  $ f_{A + A^{\top}}(\lambda):=\det (\lambda I - A-A^{\top}) $. It happens that this polynomial is also responsible for the order of the distance equation.

\begin{theorem} \label{ThDegree} One has 
\begin{equation}
\mathcal F(z)\equiv 4^n \left[ \mathcal D_{\lambda} (f_{A+A^{\top}}(\lambda)) \right]^2 z^{n(n-1)} + \mbox{lower order terms in }\ z \, .
\label{FzExp}
\end{equation}    
\end{theorem}

\textbf{Proof.} Let $ \{\mu_1,\dots, \mu_n \}$ be the spectrum of the matrix $ A+A^{\top}$ while $ P\in \mathbb R^{n\times n} $ be an orthogonal matrix reducing it to the diagonal form: 
$$
P^{\top}(A+A^{\top}) P= \diag (\mu_1,\dots, \mu_n) \,  .
$$
Apply the same transformation to the determinant (\ref{eqPhi00}):
$$
\Phi(\lambda,z) \equiv \det \left[
P^{\top}AA^{\top}P + \diag (\lambda^2-\mu_1 \lambda-z,\dots,
\lambda^2-\mu_n \lambda-z) \, .
\right]
$$
The leading term  of the polynomial $ \mathcal D_{\lambda} (\Phi(\lambda,z)) $ with respect to $ z $ coincide with that of 
$$ \mathcal D_{\lambda} \left(\prod_{j=1}^n (\lambda^2-\mu_j \lambda-z)\right) \, .
$$
The set of zeros of the polynomial under the discriminant sign is as follows
$$
\left\{ \frac{1}{2} \left(\mu_j \pm \sqrt{\mu_j^2+4z} \right) \right\}_{j=1}^n \, .
$$
Using the definition (\ref{DiscDef}) of the discriminant, one gets 
$$ \mathcal D_{\lambda} \left(\prod_{j=1}^n (\lambda^2-\mu_j \lambda-z)\right) = 
\prod_{j=1}^n (4\,z + \mu_j^2) 
\prod_{1\le j < k \le n} \left[ z^{2} (\mu_k-\mu_j)^{4} \right] \, .
$$
Coefficient of the monomial $ z^{n^2} $ in the right-hand side can be recognized, via (\ref{DiscDef}),  as the square of the  discriminant of the characteristic polynomial of $ A+ A^{\top} $. \qed

As for the determining the structure of the free term of $\mathcal F(z)$, our successes are restricted to the following 

{\bf Hypothesis.} If computed symbolically with respect to  the entries of $ A $, $ \mathcal F(0)$ has a factor $ \left[\mathcal D_{\lambda} (f_A(\lambda)) \right]^2  $.

According to Theorem \ref{ThDiscrHankel}, the polynomial $\mathcal F(z) $ can be constructed in the form of determinant of a suitable Hankel matrix.
For this aim, compute first the Newton sums $ \{s_j(z)\}_{j=0}^{4n-2} $ for  the polynomial $ \Phi(\lambda,z) $ treated with respect to $ \lambda $. Direct utilization of the formulas (\ref{Newton_sums}) requires the canonical representation (\ref{Phi_can}) for the polynomial $ \Phi(\lambda,z) $ while initially we have just only its representation in the determinantal form (\ref{eqPhi}). Fortunately, the Newton sums can be computed in an alternative way. 
Indeed, 
$$ \Phi(\lambda,z) \equiv \det (\lambda I_{2n}- W) \quad \mbox{where} \ 
W:=\left[ \begin{array}{cc} 
A^{\top} & \sqrt{z} I_n \\
\sqrt{z} I_n & A 
\end{array}
\right]
$$
and it is known that the Newton sums of the characteristic polynomial of a matrix can be computed as the traces of matrix powers:
$$
s_j(z) \equiv \tr(W^j) \quad \mbox{for} \ j \in \{0,1,\dots\} 
$$
Thus, one has
$$
s_2(z)=2 (\tr(A^2)+nz),\ s_3(z)=2(\tr(A^3)+3\, z\,  
\tr(A)),\ \dots 
$$
Compose the Hankel matrix
$$
S(z):=\left[ s_{j+k}(z) \right]_{j,k=0}^{2n-1} 
$$
and compute the sequence of its leading principal minors $ S_1(z),\dots,S_{2n}(z) $. Due to 
(\ref{SnDisc}) and (\ref{eqFz}),  
\begin{equation}
S_{2n}(z)\equiv \mathcal D_{\lambda}(\Phi(\lambda,z))\equiv \mathcal F(z) z^n \, .
\label{S2nFz}
\end{equation}
Evidently, the polynomial $ \Phi(\lambda,0) $ possesses only $ n $ double zeros, and they all are distinct provided that $ A \not\in \mathbb D $. Consequently, due to Theorem \ref{ThDiscrHankel}, one  has
$ S_{n+1}(0)=0,\dots, S_{2n}(0)=0 $.

\begin{theorem}\label{ThRealZ} Polynomial $ \mathcal F(z) $ does not have negative zeros.
The number of its positive zeros lying within the interval
$ [0,z_0], z_0>0 $ is not less than 
\begin{equation}
 | \mathcal V(1,S_1(z_0),\dots,S_{2n}(z_0)) - \mathcal V(1,S_1(0),\dots,S_{n}(0)) | \, .
 \label{Variat1}
\end{equation} 
\end{theorem}

%\mathcal V_{z_0}:= \mathcal V(1,S_1(z_0),\dots,S_{2n}(z_0)) \, .

\textbf{Proof.} The first claim of the theorem follows from the positive definiteness of the matrix \allowbreak
$ (\lambda I - A)(\lambda I - A)^{\top}-z I $ for $ z< 0$.

By Theorem \ref{ThJacobi}, the number $ \mathcal V(1,S_1(z_0),\dots,S_{2n}(z_0)) $ equals the number of complex-conjugate pairs of zeros for the polynomial $\Phi(\lambda,z_0)$. When the parameter $ z $ varies from $ 0 $ to $ z_0$, the discriminant $ \mathcal D_{\lambda} (\Phi(\lambda,z)) $ vanishes at any value of $ z $ where a pair of real zeros of $ \Phi(\lambda,z) $ transforms to a pair complex-conjugate ones or vice versa. The discriminant vanishes at these values. \qed

Theorem \ref{ThDegree} claims that the degree of the distance equation generically equals $ n(n-1)$. One can immediately watch that for the skew-symmetric matrix $ A $ this estimation is not valid. Moreover, for this type of matrices, polynomial $ \mathcal F(z) $ vanishes identically. Some other types of matrices that permit explicit representation  
for the polynomial $ \Phi(\lambda,z) $, and, as a consequence, for the value $ d(A, \mathbb D) $, in terms of the spectrum of $ A $ can be found in \cite{Kalinina_Uteshev_CASC22}. We summarize those results in the following 

\begin{theorem} \label{ThGoodMatr} Let all the eigenvalues $\lambda_1,\dots,\lambda_n $ of $ A $ be distinct. One has:
\begin{equation}
\Phi(\lambda,z)\equiv \prod_{j=1}^n [(\lambda-c)^2-(\lambda_j-c)^2-z] \quad \mbox{for} \ A=\mbox{skew-symmetric} \ + cI_n \ ,
\label{IplSkew}
\end{equation}
where $ c \in \mathbb R $ is an arbitrary scalar;
\begin{equation}
\Phi(\lambda,z)\equiv \prod_{j=1}^n (\lambda^2-z+1-2\lambda \Re(\lambda_j)) \quad \mbox{for orthogonal} \ A \  \ ;
\label{Orth}
\end{equation}
\begin{equation}
\Phi(\lambda,z)\equiv \prod_{j=1}^n \left[(\lambda-\lambda_j)^2-z \right]  \quad \mbox{for symmetric} \ A \ .
\end{equation}
  \end{theorem}
  
For the case (\ref{IplSkew}), $ \Phi(\lambda,z) $ has a multiple zero if $ n\ge 2$. For the case (\ref{Orth}), 
 $ \Phi(\lambda,z) $ has a multiple zero if $ n\ge 3$. For the both cases, the distance $ d(A, \mathbb D)$ is attained at the continuum of matrices in $ \mathbb D $ \cite{Kalinina_Uteshev_CASC22}.
 
\begin{example} Find $d(A,\mathbb D)$ for the skew-symmetric matrix
$$
A=\left[ \begin {array}{rrrr} 
0&-4&2&-1\\ 
4&0&7&3 \\
-2&-7&0&11\\ 
1&-3&-11&0
\end {array} \right] \, .
$$
\end{example}

\textbf{Solution.} Here
$$
\Phi(\lambda,z)\equiv \left( {\lambda}^{4}-2\,{\lambda}^{2}z+200\,{\lambda}^{2}+{z}^{2}-200
\,z+3249 \right)^{2} \, ,
$$
and $\mathcal D_{\lambda} (\Phi(\lambda,z)) \equiv 0$. However, if we take 
$$
\mathcal D_{\lambda} (\sqrt{\Phi(\lambda,z)})=
\mathcal D_{\lambda}({\lambda}^{4}-2\,{\lambda}^{2}z+200\,{\lambda}^{2}+{z}^{2}-200
\,z+3249)
$$
the result is the true distance equation
$$
11667456256\,z^2-2333491251200\,z+37907565375744=0 \, .
$$
Its least positive zero equals 
$$ 100-\sqrt{6751} =\frac{1}{4}(\sqrt{314}-\sqrt{86})^2  
$$ 
where $ \pm 1/2 \mathbf i (\sqrt{314}-\sqrt{86})  $ are the eigenvalues of $ A $. \qed

{\bf Remark}. Similar trick works also for the case of orthogonal matrices.

\section{Examples and Computational Aspects}
\label{SExamples}

\setcounter{equation}{0}
\setcounter{theorem}{0}
\setcounter{example}{0}
\setcounter{cor}{0}

Once the canonical form of the distance equation is computed, Wilkinson's problem is nearly solved.
Indeed, for a univariate algebraic equation, the  exact number of real zeros, as well as their location, could be trustworthy determined via purely algebraic procedures. 

{\bf Remark.} Theorem \ref{ThDegree} claims that generically the degree of the distance equation equals $n(n-1)$. The both examples below fall into this genericity. For instance,  one has $\deg \mathcal F(z)=870 $ 
for $ n=30 $.

\begin{example}   Find $d(F_n,\mathbb D) $ for  Frank's matrix \cite{Frank}
\begin{equation}
F_n=\left[
\begin{array}{cccccc}
n & n-1 & n-2 & \dots & 2 & 1 \\
n-1 & n-1 & n-2 & \dots & 2 & 1 \\
0 & n-2 & n-2 & \dots & 2 & 1 \\
0 & 0   & n-3 & \dots & 2 & 1\\
\vdots & \vdots  & & \ddots & \vdots  & \vdots \\
0 & 0 & 0 & \dots & 1 & 1
\end{array}
\right] \, .
\label{FrankM}
\end{equation}
\end{example}

\textbf{Solution.} For $ n=3$, one has
$$ \Phi(\lambda,z)=
{\lambda}^{6}-12\,{\lambda}^{5}+ \left( -3\,z+48 \right) {\lambda}^{4}
+ \left( 24\,z-74 \right) {\lambda}^{3}
$$
$$
+ \left( 3\,{z}^{2}-73\,z+48
 \right) {\lambda}^{2}+ \left( -12\,{z}^{2}+70\,z-12 \right) \lambda-{
z}^{3}+25\,{z}^{2}-33\,z+1
$$
and
$$
\mathcal F(z)=
23839360000\,{z}^{6}-476315200000\,{z}^{5}+3522206312000\,{z}^{4}-
11668368222400\,{z}^{3}
$$
$$
+16297635326400\,{z}^{2}-6895772352000\,z+
230443315200 \, .
$$
Distance equation has only real zeros, namely
$$
z_1 \approx 0.036482, \ z_2 \approx 0.648383,\ z_3 \approx 2.316991, \ 
z_4 \approx 4.954165,\ z_5 \approx 5.274176, \ z_6 = 27/4=6.75 \, .
$$
Thus, $ d(F_3,\mathbb D)=\sqrt{z_1} \approx 0.191004 $. To find the corresponding perturbation via (\ref{min_perturb}), first evaluate the multiple zero for $ \Phi(\lambda,z_1) $ via (\ref{double_zero}):
$$
\lambda_{\ast} \approx 0.602966 \, .
$$
Then evaluate the unit left singular vector of the matrix $ \lambda_{\ast} I - A $ corresponding to $ \sqrt{z_1} $:
$$
U_{\ast} \approx \left[0.639244,\, -0.751157,\, -0.164708\right]^{\top}
$$
Finally, 
$$
E_{\ast}\approx \left[\begin{array}{rrr}
-0.019161 & -0.041159 & 0.113343\\
0.022516 & 0.048365 & -0.133186 \\
0.004937 &  0.010605 & -0.029204
\end{array}
\right] \, .
$$
The nearest to $F_3$ matrix in $\mathbb D$ 
$$
B_{\ast}=F_3+E_{\ast}\approx
 \left[\begin{array}{rrr}
 2.980838 & 1.958840 & 1.113343\\
 2.022516 & 2.048365 & 0.866813\\
 0.004937 & 1.010605 & 0.970795
\end{array}
\right]
$$
possesses the spectrum $\{ \lambda_{\ast}, \lambda_{\ast}, 6-2\lambda_{\ast} \approx 4.794067 \}$.

  For $n>3$, the set of nonreal zeros for the distance equation becomes nonempty, and its cardinality, relative to that of real, increases fastly with $ n $.
\begin{center}
\renewcommand{\arraystretch}{1.2}
\begin{tabular}{ c | c | c | c | c }
$ n $  & $ d(F_n,\mathbb D)  \approx $   & coefficient size  & number of real zeros & timing (s) \\
 \hline
 $ 5 $ & $ 4.499950 \ \times 10^{-3} $ & $\sim 10^{50} $ &  $ 12 $ & $ - $ \\
 $ 10 $ & $ 3.925527 \times 10^{-8} $ & $\sim 10^{300} $ & $ 30 $ & $ - $ \\
 $ 12 $ & $ 1.849890 \times 10^{-10} $ & $\sim 10^{480} $ & $ 34 $ & $0.13 $ \\
  $ 20 $ & $ 3.757912 \times 10^{-21} $ & $\sim 10^{1690} $ & $ 62 $  &  $ 5 $ \\
 $ 30 $ & $ 1.638008 \times 10^{-36} $ & $\sim 10^{4450} $ & $ 102 $  &  $ 30 $ 
\end{tabular}.
\end{center}
The results for $F_{10}$ and $ F_{12} $ confirm estimations $d_{10}\approx 3.93\cdot10^{-8}$ and $d_{12}\approx 1.85\cdot 10^{-10}$ given in~\cite{AlamBora}. \qed

\begin{example} Find $d(K_n,\mathbb D) $ for Kahan's matrix \cite{AlamBora,Kahan}
$$
K_n=\left[
\begin{array}{cccccc}
1 & -c & -c & \dots & -c & -c \\
0 & s & -sc & \dots & -sc & -sc \\
0 & 0 & s^2 & \dots & -s^2c & -s^2c \\
 &   & \ddots & \dots &  & \\
0 & 0  & 0 & \ddots & s^{n-2}  & -s^{n-2}c \\
0 & 0 & 0 & \dots & 0 & s^{n-1}
\end{array}
\right] \quad \mbox{for} \ s^2+c^2=1 \, .
$$
\end{example}

{\bf Solution.} We present computational results for two specialization of parameter values. The first one is $ s= 3/5, c=4/5 $:

\begin{center}
\renewcommand{\arraystretch}{1.2}
\begin{tabular}{ c | c | c | c | c }
$ n $  & $ d(K_n,\mathbb D)  \approx $  & 
coefficient size  & number of real zeros 
& timing (s) \\
 \hline
 $ 5 $ & $ 1.370032 \times 10^{-3}  $ &  $\sim 10^{310} $ &  $ 8 $ & $ - $ \\
 $ 10 $ & $ 5.470834 \times 10^{-6} $ & $\sim 10^{2970} $ & $ 48 $ & $ - $ \\
 $ 15 $ & $ 2.246949 \times 10^{-8}  $ &  $\sim 10^{10590} $  & $ 138 $ & $6.7 $ \\
  $ 20 $ & $ 9.245309 \times 10^{-11} $ & $\sim 10^{25730} $  & $ 288 $  &  $ 145.4 $ \\
 $ 25 $ & $  3.984992\times 10^{-10} $ & $\sim 10^{52910} $   & $ 258 $  &  $ 218.23 $  \\
 $ 30 $ & $ 1.240748\times 10^{-11} $ & $\sim 10^{92460} $ & $ 464 $ & $937.66$
 
\end{tabular}
\end{center}

The second test series correspond to a specialization $s^{n-1}=1/10 $ treated in~\cite{AlamBora}. For this case, an extra difficulty results from approximation of the entries of the matrix $ K_n $ as rational numbers. This results in increasing the length of the coefficients of the distance equation. Compared with the previous case, the timing increases drastically, i.e. more than $ 10^2 $ times for the same specializations of $ n $.   

\begin{center}
\renewcommand{\arraystretch}{1.2}
\begin{tabular}{ c | c | c }
$ n $  & $ d(K_n,\mathbb D)  \approx $   & number of real zeros  \\
 \hline
 $ 6 $ & $ 4.704940 \times 10^{-4}  $ & $ 10 $  \\
 $ 10 $ & $ 1.538157 \times 10^{-5} $ & $ 18 $  \\
 $ 15 $ & $ 4.484974 \times 10^{-7}  $ &  $ 28 $  \\
  $ 20 $ & $ 1.904858 \times 10^{-8} $ &  $ 38 $  \\
%$ 25 $ & $ 1.638008 \times 10^{-36} $ &  $ 102 $   
\end{tabular}
\end{center}
The results for $K_{6},K_{15}$ and $K_{20}$ confirm estimations  given in~\cite{AkFrSp}. \qed

It should be emphasized however that computation of the whole sets of real zeros for the distance equation is redundant for evaluation of $d(A,\mathbb D) $. We need to find just only  the least positive zero of $ \mathcal F(z)$. For this aim, the determinantal representation (\ref{S2nFz}) for this polynomial might be sufficient for the real zero localization. According to Theorem \ref{ThRealZ}, the lower estimate for the number of real zeros of $ \mathcal F(z) $ lying within the interval $[0,z_0], z_0 >0$ is given by the number (\ref{Variat1}). If this number is not zero then at least one real zero for $ \mathcal F(z) $ lies in $ [0,z_0]$, and the next step in its localization might be the treatment of the matrix  $ S(z_0/2) $. 

Experiments with the Frank's matrix (\ref{FrankM}) demonstrate 
the unambiguity of the zero isolation process. For the matrix $ F_{10}$, one has
$\mathcal V(1,S_1(0),\dots,S_{10}(0))=0 $, i.e. all the eigenvalues of $ A $ are real. Then (\ref{Variat1}) coincides with 
$$ \mathcal V_{z_0}:= \mathcal V(1,S_1(z_0),\dots,S_{10}(z_0),\dots, S_{20}(z_0)) \, . $$
Some specializations for $ z_0 $
\begin{center}
\begin{tabular}{ c || c | c | c | c }
$ z_0 $  & $ 10^{-3} $  &  $ 10^{-9} $ & $ 2\times 10^{-15} $
& $ 10^{-15} $ \\
\hline
$ \mathcal V_{z_0} $ & 5  & 3 & 1 & 0 
\end{tabular}
\end{center}
demonstrate that the number of real zeros of $ \mathcal F(z) $ lying in any interval $ [0,z_0]   $ happens to be equal to $ \mathcal V_{z_0} $. For instance, there are precisely 
$ 5 $ zeros within the interval $[0,10^{-3}]$, namely
$$
1.540976\times 10^{-15},\  7.739368\times  10^{-15}, 7.463686 \times 10^{-13},\  1.403045 \times 10^{-9}, 1.412301 \times 10^{-5} \, .
$$
However, for the case of the matrix
$$
\left[ \begin {array}{rrr} 1&1&-2\\ 2&1&0
\\ -3&1&1\end {array} \right] 
$$
variations $  \mathcal V_{0.4}=0,  \mathcal V_{0.5}=1,  \mathcal V_{2.25}=0  $ permit one to locate single zeros within the intervals $[0.4, 0.5]$ and $[0.5,2.25] $ but are unable to detect this number for $[0.4, 2.25]$.

\section{Counterexamples}
\label{SExcep}

\setcounter{equation}{0}
\setcounter{theorem}{0}
\setcounter{example}{0}

We exemplify here two cases
\begin{itemize}
\item[(a)]  The minimal positive zero of the distance equation not always provides the value $d^2(A,\mathbb D)$ even if we restrict ourselves to the rank $ 1 $ perturbation matrices;
\item[(b)] The distance $d(A,\mathbb D)$ is not always provided by the rank $ 1 $ perturbations.
\end{itemize}

\begin{example} For the matrix
$$
A(\epsilon)=
\left[ \begin {array}{cccc} 0&1&1&0\\ -1&0&0&1
\\ \epsilon&0&0&1\\ 0&0&-1&0
\end {array} \right] \, ,
$$
find $ d(A(\epsilon),\mathbb D) $ for $ \epsilon >0$.
\end{example}

\textbf{Solution.} Distance equation is provided by the polynomial
$$
\mathcal F(z)\equiv 65536\epsilon^8 \left[(\epsilon+2)^4 z^2 -2\epsilon(\epsilon+8)(\epsilon+2)^2 z + \epsilon^2(\epsilon-8)^2\right]^2
\cdot \left[(\epsilon+1)z-3\epsilon-1\right]^4 
$$
$$
\times (z^2-3\,z+1)  
\left[{z}^{2}-\left( {\epsilon}^{2}+3 \right) z+({\epsilon}+1)^2\right] \, .
$$
Its zeros are
$$
z_1=\frac{\epsilon(\sqrt{\epsilon}-\sqrt{8})^2}{(\epsilon+2)^2},\ z_2=\frac{\epsilon(\sqrt{\epsilon}+\sqrt{8})^2}{(\epsilon+2)^2}, z_3=\frac{3\epsilon+1}{\epsilon+1}, 
$$
$$
z_4=\frac{3-\sqrt{5}}{2} \approx 0.381966, \ z_5=\frac{3+\sqrt{5}}{2} \approx 2.618033 \, ,
$$
$$
z_6=\frac{1}2\left({\epsilon}^{2}+3-|\epsilon-1| \sqrt {{\epsilon}^{2}+2\,{\epsilon}^{2}+5}\right),\ z_7=
\frac{1}2\left({\epsilon}^{2}+3+|\epsilon-1| \sqrt {{\epsilon}^{2}+2\,{\epsilon}^{2}+5}\right)
$$
are all real. Zero $z_4$ is simple, it coincides with the square of a singular value of the matrix $ A $, and the polynomial $\Phi(\lambda,z_4) $ has the real double zero $\lambda_4=0$. The corresponding value of the distance function from $ A $ to $\mathbb D$ does not depend on $\epsilon$, it equals\footnote{Amazing coincidence with the reciprocal to the \textit{golden ratio}!}
$$ 
\sqrt{z_4}=\frac{\sqrt{5}-1}{2} \approx 0.618033 \, .
$$
The corresponding perturbation and matrix in $\mathbb D$ are as follows:
$$
E_4=\frac{1}{10} \left[\begin{array}{cccc}
0 & \sqrt{5}-5 & (3\sqrt{5}-5 & 0 \\
0 & 0 & 0 & 0 \\
0 & 0 & 0 & 0 \\
0 & -2 \sqrt{5} & 5-\sqrt{5} & 0
\end{array}
\right], B_4=\frac{1}{10}
\left[\begin{array}{cccc}
0 & 5+\sqrt{5} & 5+3\sqrt{5} & 0\\
-10 & 0 & 0 & 10\\
10\epsilon & 0 & 0 & 10\\
0 & -2\sqrt{5}& -5-\sqrt{5} & 0 
\end{array}
\right] \, .
$$
Double eigenvalue of $B_4$ is just $0$.

Next, we do not need to treat the zeros $z_6, z_7$ and $z_3$, since they are greater than $ z_4$. Also $ z_2 >z_1$, therefore, the two zeros that can compete for the distance value are $z_1$ and $z_4$.  It can be verified that 
$$
z_1 \le z_4  \ \mbox{iff}\  \epsilon \le \epsilon_{2}
\ \mbox{where} \ \epsilon_{2} = 2\sqrt{2}(\sqrt{5}+3)\sqrt{\sqrt{5}+2}+7\sqrt{5}+15\approx 61.133652 \, .
$$
It looks like $ d(A,\mathbb D)=\sqrt{z_1} $ for $\epsilon \le \epsilon_{2} $. However, this is not true for some subinterval in $[0,\epsilon_{2}]$. Indeed, $ z_1$ is a double zero for $\mathcal F(z)$, and polynomial  $\Phi(\lambda,z_1) $ possesses two double zeros:
$$
\lambda_{1,2}=\pm\frac{\sqrt{K(\epsilon)}}{\epsilon+2} \quad \mbox{where} \ K(\epsilon):=\sqrt{2}(\epsilon-\sqrt{2}\sqrt{\epsilon}+2)\left(\sqrt{\epsilon}+\frac{1}{\sqrt{2}}\right)\left(\sqrt{\epsilon}+\frac{\sqrt{5}+1}{\sqrt{2}}\right)\left(\sqrt{\epsilon}-\frac{\sqrt{5}-1}{\sqrt{2}}\right) \, .
$$
These zeros are real only for  
$$\epsilon \ge \epsilon_1\  \mbox{where} \  \epsilon_1:=3-\sqrt{5} \approx 0.763932 \, . $$ 
For the values $ \epsilon < \epsilon_2 $, the minimal positive zero of the distance equation is not responsible for the distance from $ A $ to $\mathbb D$.

It seems that $ d(A,\mathbb D)=\sqrt{z_4} $ for $\epsilon < \epsilon_2$. However, this statement is also invalid for some subinterval of the parameter values. The matrix
$$
\widetilde{E}(\epsilon):=
\frac{\epsilon (8-\epsilon)}{(\epsilon^2+16)^2}
\left[ \begin {array}{cccc} 0&-4\,\epsilon&{\epsilon}^{2}&0
\\ -4\,\epsilon&0&0&{\epsilon}^{2}
\\ -16&0&0&4\,\epsilon\\ 0&-16&4\,
\epsilon&0\end {array} \right]
$$
represents a rank $ 2 $ perturbation that provides for the matrix $ A(\epsilon) + \widetilde{E}(\epsilon) $ a pair of double eigenvalues 
$$\lambda_{1,2} =\pm\frac{1}{\epsilon^2+16}
\sqrt{\left( {\epsilon}^{2}+4\,\epsilon-16 \right)  \left( 3\,{\epsilon}^{
2}+4\,\epsilon+16 \right)} \,  .
$$
These eigenvalues are non-real for $\epsilon < 2(\sqrt{5}-1) \approx 2.472136 $.  For these parameter values, one has
$$
\| \widetilde{E}(\epsilon) \|=\frac{\sqrt{2}\epsilon (8-\epsilon)}{\epsilon^2+16} 
$$
and this value is lesser than $ \sqrt{z_1}$ for $ \epsilon < \epsilon_{c} $ where $ \epsilon_{c} $ denotes the least positive zero of the polynomial 
$$
{\epsilon}^{8}-80\,{\epsilon}^{7}-368\,{\epsilon}^{6}-1024\,{\epsilon}
^{5}+64\,{\epsilon}^{4}-9216\,{\epsilon}^{3}-16384\,{\epsilon}^{2}-
32768\,\epsilon+65536 \, ;
$$
i.e. $ \epsilon_c \approx 1.055249 $.

\begin{center}
\begin{minipage}[t]{125mm}
%\graphicspath{{Illustrations/}}
\includegraphics[width=100mm]{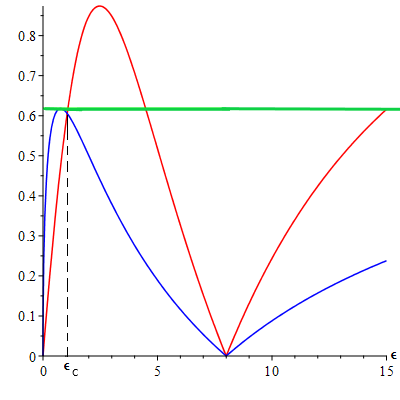}
\end{minipage}
\end{center}

\begin{center}
Figure~1.  
\end{center}

Summarizing:
$$
d(A(\epsilon), \mathbb D)=
\left\{ \begin{array}{cl}
\sqrt{2}\epsilon (8-\epsilon)/(\epsilon^2+16)   & \mbox{if}\  \epsilon \in [0, \epsilon_c]\\
\sqrt{\epsilon}|\sqrt{\epsilon}-\sqrt{8}|/(\epsilon+2) & \mbox{if} \ \epsilon \in [ \epsilon_c, \epsilon_2] \\
(\sqrt{5}-1)/2  & \mbox{if} \
 \epsilon > \epsilon_2
\end{array}
\right.
$$
The plot is displayed in Fig. 1 (the first formula --- red, the second one --- blue, the third one --- green). \qed

\textbf{Remark.} As it is mentioned in Introduction, the case where $d(A,\mathbb D)$ is achieved at the rank $2$ matrix (i.e. the nearest in $ \mathbb D $ matrix possesses two double imaginary eigenvalues) is beyond our consideration. We are not able even to conjecture whether this is a zero probability event or not.

\section{Complex Perturbations}
\label{GenCase}

\setcounter{equation}{0}
\setcounter{theorem}{0}
\setcounter{example}{0}

The method proposed above can be extended to the case of complex perturbations. For a real matrix $A$, we are now looking for the distance to the nearest complex matrix $ B $ with multiple eigenvalue:
$$
d_C(A,\mathbb D):=\min \|B-A\| \quad \mbox{subject to} \ \mathcal D_{\lambda}(f_B(\lambda))=0,\ B\in \mathbb C^{n\times n} \, .
$$

\textbf{Warning.} The present section should be considered as a draft of a separate publication to be prepared sometime afterwards. We skip here the details of algebraic backgrounds, proofs of theoretical results and do not bother ourselves with mentioning that the validity of some of the declared results is subject to several extra assumptions preventing the appearance of  
troubles similar to those dealt with in the previous section.

Consider the polynomial
\begin{equation}
\Theta(a,b,z)=\det\left[\left((a+b \mathbf i)I-A\right)\left((a-b\mathbf i)I-A^{\top}\right)-zI\right] \, 
\end{equation}
and generate the system of algebraic equations
\begin{equation}
 \Theta=0,\ \partial\Theta /\partial
a =0, \ \partial\Theta/ \partial b=0 \, .
\label{compl1}
\end{equation}
We are looking for the real solutions to this system.
Since 
$$ \Theta(a,0,z)  \stackrel{(\ref{eqPhi})}\equiv \Phi(a,z) \, ,$$ 
this solution set includes that for the system (\ref{eqsysT}).

\begin{theorem} \label{ThEven} If the system (\ref{compl1}) possesses a solution $(a_0,b_0,z_0)$ with $ b_0\ne 0 $ then it has the solution 
$(a_0,-b_0,z_0)$.
\end{theorem}

\textbf{Proof. }
Polynomial $ \Theta(a,b,z) $ is even in $ b $:
$$
\Theta(a,-b,z)=\det \left[ ((a+\mathbf i b)I-A^{\top})((a-\mathbf i b)I-A) - zI \right]
$$
$$
=\det \left[ \left\{((a+\mathbf i b)I-A^{\top})((a-\mathbf i b)I-A)\right\}^{\top} - zI \right]
$$
$$
=\det \left[((a-\mathbf i b)I-A^{\top}) ((a+\mathbf i b)I-A)  - zI \right]=\Theta(a,b,z) \, .
$$
Consequently $ \Theta^{\prime}_a $ is even in $ b $ while $ \Theta^{\prime}_b $  is odd $ b $. The latter becomes even on dividing by $ b $. \qed

Our aim is to eliminate the variables $a $ and $ b $ from the system (\ref{compl1}), i.e. to find the \textbf{bivariate discriminant }
$\mathcal D_{a,b}(\Theta)$ for the polynomial $ \Theta(a,b,z) $ treated with respect to these variables. 

The discriminant $\mathcal D_{x,y}(F)$  of a polynomial $ F(x,y,z) \in \mathbb C[x,y,z] $ is formally defined as the result of elimination of variables $x$ and $ y $ from the system of equations
\begin{equation}
F=0,\ \partial F / \partial x=0, \  \partial F / \partial y=0 \, . 
\label{EqforD}
\end{equation}
This is a polynomial in $ z $ and its vanishment at $ z=z_0 \in \mathbb C $ is the necessary and sufficient condition for the existence of solution $(x_0,y_0,z_0) \in \mathbb C^3 $ to the system (\ref{EqforD}), or equivalently, for the existence of the multiple zero $(x_0,y_0) $ for the polynomial $  F(x,y,z_0) $. Constructive computation of discriminant 
can be implemented in several ways, and we will exemplify below the procedure based of the Bézout construction of the resultant \cite{BikkerUteshev}.

\begin{theorem} The discriminant $ \mathcal D_{a,b}(\Theta(a,b,z)) $  is factorized as follows:
\begin{equation}
\mathcal D_{a,b}(\Theta(a,b,z))\equiv z^{n(n+1)/2} \mathcal F(z) \widetilde{\mathcal F}(z)   \, . 
\label{Discrim2Ide}
\end{equation}
Here  $ \mathcal F(z) $ is defined by (\ref{eqFz}), while 
$$
\widetilde{\mathcal F}(z) \in \mathbb R[z], \ \deg \widetilde{\mathcal F}(z) =n(n-1)(n-2)/2 \, ,
$$
(For $ n=2 $ polynomial $ \widetilde{\mathcal F}(z) $ is just a constant).
\end{theorem}

According to Section \ref{SDisEq}, the distance equation $ \mathcal F(z)=0$ is responsible for the rank $ 1 $ real perturbation that provides the distance $d(A,\mathbb D) $.
It turns out that the equation
$$
 \widetilde{\mathcal F}(z) = 0
 $$ 
 is responsible for the rank $ 1 $ imaginary perturbation. Its real zero $ \widetilde{z}_0 $ corresponds to a pair of multiple zeros 
of the polynomial $ \Theta(a,b, \widetilde{z}_0) $, and these zeros are either in the form  $ (a_0, \pm \beta_0) $ or in the form
$ (a_0, \pm \mathbf i \beta_0) $ with real $ \beta_0 $. We are definitely interested only in the real solutions for the 
system (\ref{compl1}).

\begin{theorem}
\label{Thcomplex} Let the system (\ref{compl1}) possess a real solution $(a_0,b_0,\widetilde{z}_0)$ with $\widetilde{z}_0 >0,b_0\ne 0$.
Denote $ U_{0} \in \mathbb C^n, \| U_{0} \|=1 $ the left singular vector for the matrix 
$(a_0+\mathbf i b_0)I-A$ corresponding to the singular value 
$\sqrt{\widetilde{z}_0}$.
Then  the rank $ 1 $ perturbation
\begin{equation}
E_{0}=U_{0} U_{0}^{\mathsf{H}} ((a_0+\mathbf i b_0) I-A)
\label{Ecomplex}
\end{equation}
is such that $\| E_{0} \|= \sqrt{\widetilde{z}_0} $ and the matrix $ B_0=A+E_0 \in \mathbb C^{n\times n}$ possesses the double eigenvalue $ a_0+\mathbf i b_0 $.
\end{theorem}

\textbf{Remark.} Evidently, the matrix $\overline{E_0} $ provides for the matrix 
$ \overline{B_0}=A+\overline{E_0} $ the double
eigenvalue $ a_0-\mathbf i b_0 $.

In view of Theorem \ref{Thcomplex}, the distance $ d_C(A,\mathbb D) $ results from the competition between the least positive zero of $ \mathcal F(z)$ and that minimal positive zero of 
$ \widetilde{\mathcal F}(z) $ that corresponds to the real solution for the system (\ref{compl1}).

Computation of the polynomial $ \widetilde{\mathcal F}(z) $ can be simplified if we take into account Theorem \ref{ThEven}. Substitute 
$$ \mathfrak b:=b^2 $$ 
in the polynomials of the system (\ref{compl1}) and denote
$$
\Xi(a,\mathfrak b,z):=\Theta(a,b,z), \  \Xi_a(a,\mathfrak b,z):=\Theta^{\prime}_a(a,b,z), \ 
\Xi_{\mathfrak b}(a,\mathfrak b,z):=\Theta^{\prime}_b(a,b,z)/b  \, .
$$

\begin{theorem} The result of elimination of variables $a $ and $ \mathfrak b $ from the system
\begin{equation}
\Xi=0,\ \Xi_a=0, \Xi_{\mathfrak b}=0 
\label{Eq_Xi}
\end{equation}
is the equation 
$$
z^{n(n-1)/2} \widetilde{\mathcal F}(z)=0 \, .
$$
\end{theorem}

If $ \widetilde z_0 $ is a positive zero of $ \widetilde{\mathcal F}(z) $, the corresponding real solution to the system (\ref{Eq_Xi}) might have the $ \mathfrak b $-component either positive or negative. We are interested only in the positive variant.

\begin{example} Find $ d_C(A,\mathbb D)$ for
$$
A= \left[ \begin {array}{rrr} 0&1&0\\ 0&0&1
\\ -91&-55&-13
\end {array} \right] \, .
$$
\end{example}

\textbf{Solution.} First compute the polynomial  $ \mathcal F(z) $ via (\ref{eqFz}):
$$
\mathcal F(z) := 33076090700402342058246544\,z^6-377039198861306289080145178864\, z^5
$$
$$
+937864902703881321034450183916\,z^4-771868276098720970149792503999\,z^3
$$
$$
+211070978787821517684022650624\,z^2
$$
$$
-510584100140452518540394496\,z+319295875259784560640000 \, .
$$
Its real zeros  are as follows
$$
z_1\approx 0.739336,\  0.765571,\ 0.980468,\  11396.658548 \, .
$$
Next compose the polynomial $ \Xi(a,\mathfrak b,z) $:
$$\Xi(a,\mathfrak b,z)=-z^3+(3a^2+3\mathfrak b+26a+11477)z^2$$
$$-(3\,a^4+6\,a^2\mathfrak b+3\,\mathfrak b^2+52a^3+52a\mathfrak b+11756a^2+11536\mathfrak b+11466\,a+19757)z
$$
$$
+\left( {a}^{2}+\mathfrak b+14\,a+49 \right)  \left( ({a}^{2}+\mathfrak b+6\,a+13)^2-
16\,\mathfrak b \right) \, .
$$
Now we trace briefly the procedure  of elimination of $ a $ and $ \mathfrak b $ from the system (\ref{Eq_Xi}). Consider the monomial sequence
$$
\mathbb M:=\{\mathfrak m_{j}(a,\mathfrak b)\}=\{1,a,\mathfrak b, \mathfrak b^2\} \, .
$$
It is possible \textbf{to reduce} the polynomial $ \mathfrak m_j  \Xi $ \textbf{modulo} $ \Xi_a $ and $ \Xi_{\mathfrak b} $, i.e. to 
find the polynomials $ \{\beta_{jk}(z) \}_{j,k=1}^4 \subset \mathbb R[z] $ and $ \{p_{11}(a,\mathfrak b,z), p_{j2}(a,\mathfrak b,z)\}_{j=1}^4 \subset \mathbb R[a,\mathfrak b,z] $ satisfying the identity
$$
 m_j  \Xi \equiv \beta_{j1}(z)+ \beta_{j2}(z) a + \beta_{j3}(z) \mathfrak b + \beta_{j4}(z) \mathfrak b^2 +  p_{j1} 
  \Xi_a  + p_{j2}    \Xi_{\mathfrak b}   \ \mbox{for} \ j\in \{1,2,3,4\} \, .
$$
For instance, 
$$
\beta_{11}(z)= -17718805921\,z^2+610367232\,z+22937600, \ \beta_{12}(z)= -39353600\, z+5324800, \ 
$$
$$
\beta_{13}(z)=146694400\,z-512000, \beta_{14}(z)=-307200, \dots, 
$$
$$
\beta_{44}(z)=
-{\scriptstyle 76550493273549926400}\,z^3+{\scriptstyle 162810741053705011200}\,z^2-{\scriptstyle 1867736871075840000}\,z-{\scriptstyle 50331648000000} \, .
$$
Compose the Bézout matrix 
$$
\mathfrak B(z):= \left[ \beta_{jk}(z) \right]_{j,k=1}^4 \, .
$$
Then
$$
\det \mathfrak B(z) \equiv  z^3\widetilde{\mathcal F}(z)
$$
where
$$
\widetilde{\mathcal F}(z) =  412324266119803814719539025\,{z}^{3}+
33923334498676415590177600\,{z}^{2}
$$
$$
+691077589890510378371072\,z-
899669298077697638400 \, . 
$$
For any zero $ \widetilde z_0 $ of this polynomial, the corresponding $a$  and $ \mathfrak b $ components of the solution to the system (\ref{Eq_Xi}) can be obtained in the following way. Denote by $ \{\mathfrak B_{4j}\}_{J=1}^4 $ the cofactors of 
$ \det \mathfrak B $ corresponding to the entries of the last row of the matrix $ \mathfrak B $. Then the $ a $-component of solution is connected with the $ z $-component as
$$
a=\frac{\mathfrak B_{42}}{\mathfrak B_{41}}= \frac{43719663040898080379\,{z}^{2}+2929017747573439808\,z+
29336262189312000}{2(624300876564482975\,z^2-226254560538037856\,z-3469512291865600)}
$$
while the $\mathfrak b $-component as
$$
\mathfrak b=\frac{\mathfrak B_{43}}{\mathfrak B_{41}}= \frac{{\scriptstyle 3083432482762007609519}\,{z}^{3}+{\scriptstyle 1101690698089389073600}\,
{z}^{2}+{\scriptstyle 67186386329988787456}\,z-{\scriptstyle 129087561954918400}}{16(624300876564482975\,z^2-226254560538037856\,z-3469512291865600)}
$$
Polynomial $ \widetilde{\mathcal F}(z) $ possesses a single real zero, namely\footnote{All the decimals in the following approximation are error-free.}
$$
\widetilde{z}_1 \approx 0.0012268490707391199222512104943 \, ,
$$
and substitution of this value into the last formulas yields
$$ a= a_1 \approx -4.403922040624116177182912013601, \   \mathfrak b = \mathfrak b_1 \approx  0.750705046015830894563798035515 \, . 
$$
Since $ \mathfrak b_1>0 $, one may claim that  
$$
d_C(A,\mathbb{D})=\sqrt{\widetilde{z}_1}\approx 0.035026405335676681771543151648 \, .
$$
The two  perturbations in $ \mathbb C^{3\times 3} $ providing this distance correspond to the solutions
$$
(a_1,b_1,\widetilde{z}_1) \ \mbox{and} \ (a_1,-b_1,\widetilde{z}_1) \ \mbox{where} \ \ b_1=\sqrt{\mathfrak b_1} \approx 0.866432366671415902596255690462 \, .
$$
of the system (\ref{compl1}). Let us compute via (\ref{Ecomplex}) the one  corresponding to $(a_1,-b_1,\widetilde z_1) $. 
The unit left singular vector of
$(a_1-\mathbf i b_1)I-A$ corresponding to the singular value 
$\sqrt{\widetilde{z}_1}$ is as follows 
$$U_1 \approx \left[ 
 0.930609,\  
 0.360923+
 0.039918\,\mathbf i,\ 
 0.045052+
 0.008866\,\mathbf i
\right]^{\top}
$$
and the minimal perturbation
$$
E_1\approx \left[\begin{array}{ccc}
0.001289-0.000442 \mathbf i&-0.007120+0.000832 \mathbf i&0.031666+0.002551 \mathbf i\\
0.000519-0.000116 \mathbf i&-0.002797+0.000017 \mathbf i&0.012172+0.002348 \mathbf i\\
0.000067-0.000009 \mathbf i&-0.000353-0.000028 \mathbf i&0.001509+0.000425 \mathbf i
\end{array}\right]\, .
$$
The spectrum of the matrix $A+E_1$ is 
$$ \{ a_1-\mathbf i b_1, a_1-\mathbf i b_1 ,-13-2(a_1-\mathbf i b_1) \approx -4.192156-1.732865 \mathbf i \} \, . $$
\qed

To test the performability of the algorithm sketched in the present section, we chose the next  matrix from the Matlab gallery($'$grcar$'$,6).

\begin{example} Find $ d_C(A,\mathbb D)$  for
$$
A= \left[ \begin {array}{rrrrrr} 1&1&1&1&0&0\\ -1&1&1&1&1&0
\\ 0&-1&1&1&1&1\\ 0&0&-1&1&1&1\\ 0&0&0&-1&1&1\\ 0&0&0&0&-1&1
\end {array} \right] \, . 
$$
\end{example}

\textbf{Solution.} Here the minimal zero of $\mathcal F(z)$ equals
$z_1\approx 0.116565 $ 
and that of $ \widetilde{\mathcal F}(z) $ equals
$$
\widetilde{z}_1 \approx 0.04630491415327188209539627157 \, .
$$
The latter corresponds to the real solution for the system (\ref{compl1}):
$$  
(a_1,\pm b_1, \widetilde{z}_1) \  \mbox{where} \ a_1 \approx 
 0.753316, \ b_1 \approx -1.591155 \, .
$$
Thus, one obtains
$$d_C(A,\mathbb D) = \sqrt{\widetilde{z}_1}\approx 0.2151857666140395125353 \, .$$
This confirms estimation $d_C(A,\mathbb D) \approx 0.21519 $
from~\cite{AkFrSp, AlamBora}. 

For the solution $(a_1,b_1, \widetilde{z}_1) $, the spectrum of the nearest to $A$ matrix in $ \mathbb D$ is as follows
$$
\{ 0.361392-1.944783 \mathbf i,1.139422-1.239762 \mathbf i,1.502453-0.616966 \mathbf i,1.490100+0.619201 \mathbf i,a_1+\mathbf i b_1, a_1+\mathbf i b_1 \} \, .
$$

\section{Conclusion}
\label{SConclus}

We have investigated Wilkinson's problem for the distance evaluation from a given matrix to the set of matrices possessing multiple eigenvalues. The proposed approach consists in the construction of distance equation with the zero set containing the critical values of the squared distance function. This construction is realized in the ideology of symbolic computations, i.e. the procedure consists of a finite number of elementary algebraic operations on the entries of the matrix. 

The representation of the distance equation with the aid of the discriminant function should not be taken as a complete surprise. Indeed, the Wilkinson's  problem is the one of evaluation the distance to the discriminant manifold in the space of matrix entries. Hence, in view of this circumstance, the appearance of the discriminant in a solution to the problem is somehow natural. The more astonishing is the emergence of the discriminant in nearly {\it any} problem of distance evaluation from a point to an algebraic manifold in a multidimensional space \cite{UteshevCherkasov,UteshevYashina2015}.

Direction for further research is clearly related the stuff of Section \ref{SExcep}, i.e. the problem of existence the rank $2 $ minimal perturbation providing $ d(A,\mathbb D) $.

%\section{Acknowledgment}
%This research was supported by the St.\,Petersburg State University grant \textbf{9.38.674.2013}.

\end{document}